\documentclass[aps,prb,twocolumn,showpacs,superscriptaddress]{revtex4-1}
\usepackage{graphicx}

\begin{document}
\bibliographystyle{apsrev}
\title{The evolution of antiferromagnetic susceptibility to uniaxial pressure in Ba(Fe$_{1-x}$Co$_{x}$)$_2$As$_2$}

\author{Chetan Dhital}
\affiliation{Department of Physics, Boston College, Chestnut Hill, Massachusetts 02467, USA}
\author{Tom Hogan}
\affiliation{Department of Physics, Boston College, Chestnut Hill, Massachusetts 02467, USA}
\author{Z. Yamani}
\affiliation{Canadian Neutron Beam Centre, National Research Council, Chalk River, Ontario, Canada K0J 1P0}
\author{Robert J. Birgeneau}
\affiliation{Materials Science Division, Lawrence Berkeley National Laboratory, Berkeley, California, 94720, USA}
\author{W. Tian}
\affiliation{Quantum Condensed Matter Division, Oak Ridge National Laboratory, Oak Ridge, Tennessee 37831-6393, USA}
\author{M. Matsuda}
\affiliation{Quantum Condensed Matter Division, Oak Ridge National Laboratory, Oak Ridge, Tennessee 37831-6393, USA}
\author{A. S. Sefat}
\affiliation{Neutron Scattering Science Division, Oak Ridge National Laboratory, Oak Ridge, Tennessee 37831-6393, USA}
\author{Ziqiang Wang}
\affiliation{Department of Physics, Boston College, Chestnut Hill, Massachusetts 02467, USA}
\author{Stephen D. Wilson}
\email{stephen.wilson@bc.edu}
\affiliation{Department of Physics, Boston College, Chestnut Hill, Massachusetts 02467, USA}

\begin{abstract}
Neutron diffraction measurements are presented measuring the responses of both magnetic and structural order parameters of parent and lightly Co-doped Ba(Fe$_{1-x}$Co$_{x}$)$_2$As$_2$ under the application of uniaxial pressure.  We find that the uniaxial pressure induces a thermal shift in the onset of antiferromagnetic order that grows as a percentage of $T_{N}$ as Co-doping is increased and the superconducting phase is approached.  Additionally, as uniaxial pressure is increased within parent and lightly-doped Ba(Fe$_{1-x}$Co$_{x}$)$_2$As$_2$ on the first order side of the tricritical point, we observe a decoupling between the onsets of the orthorhombic structural distortion and antiferromagnetism.  Our findings place needed constraints on models exploring the nematic susceptibility of the bilayer pnictides in the tetragonal, paramagnetic regime.
\end{abstract}


\maketitle
\section{Introduction}
One of the central questions in understanding the electronic phase behavior of the iron pnictide high temperature superconductors (high-$T_c$) remains the unresolved origin of their ubiquitous tetragonal-to-orthorhombic structural distortions in both parent and underdoped concentrations\cite{kamihara2008iron,chu2009determination,canfield2009decoupling,nandi2010anomalous,Wilsonneutrondiffraction,li2009structural,wen2011materials,stewart2011superconductivity}.  While the distortion itself is subtle---resulting in a relative elongation of the basal plane $a$-axis by $\approx1\%$--- it is widely believed to be a secondary effect driven by electronic symmetry breaking such as orbital order\cite{ORBITALORDERPRL,yi2011symmetry,philips,MYi,yzhang,WLEE2009} or low energy spin fluctuations\cite{dai2012magnetism,harringer,paglione2010high,lynn2009neutron,mazinnature,fernandes2012manifestations}.  In a number of scenarios considered, the microscopic origin for this structural distortion is rooted in the presence of an otherwise hidden, electronic, nematic phase whose fluctuations are ultimately suggested to play a role within the superconducting pairing mechanism\cite{chu2012divergent,Lstojchevska,harringer,fernandes2012manifestations,liangdagotto,hu2012nematic,kasahara2012electronic,allan2013anisotropic,fernandes2013nematic}.  To date however, this scenario remains a subject of active investigation. As part of this, one of the key metrics sought as a signature of nematicity is an indication of $C_4$-symmetry breaking within the electronic properties of the iron pnictides within the nominally paramagnetic, tetragonal ($C_4$-symmetric) phase.  We emphasize that nematic long range order cannot occur if the tetragonal $C_4$ symmetry is not broken; however, there may be pronounced nematic fluctuation effects in the $C_4$-symmetric phase.

Numerous experimental probes such as dc-transport\cite{chu2012divergent,chu2010planeSCIENCE,kuomagnetoelastic,Blombergtensile,UCHIDATRANSPORT,blomberg2013sign}, optical conductivity\cite{UCHIDAOPTICAL,Duszaanisotropic,gallais2013observation,Tanataroptical}, scanning tunneling microscopy\cite{allan2013anisotropic,PASHUPATHY}, angle-resolved photoemission\cite{yi2011symmetry,MYi,yzhang} neutron scattering\cite{dhital2012effect,harringer,Yusongprb}, and magnetic torque measurements\cite{kasahara2012electronic} have either directly or indirectly resolved the presence of the electronic behavior violating the $C_4$ rotational symmetry within the FeAs planes of different families of iron pnictide high-$T_c$ systems. Initial studies relied on bulk probes of crystals which manifested twin structural domains below their tetragonal-to-orthorhombic structural distortion temperatures ($T_S$).  These bulk studies necessarily rely on a symmetry breaking field which biases twin domain formation and allows uniquely defined directions within the basal planes of these systems. The symmetry breaking fields are typically comprised of simple uniaxial strain applied to the underlying crystalline lattice; however magnetic fields\cite{chu2010plane} are also utilized---in either case, the strong spin-lattice coupling inherent to these materials necessarily results in the perturbation of both the underlying nuclear lattice and the antiferromagnetic order as the system is prepared for study.  Correspondingly, the core observation of the nematic behavior inherent to these systems, as seen via bulk probes, stems from their dramatic susceptibility to the perturbations brought on by these external symmetry breaking fields, that ultimately allow the nematic order parameter to develop.

A variety of scenarios have been proposed in modeling the microscopic origin of the nematic susceptibility in the iron-based high-$T_c$ compounds such as orbital ordering/fluctuations\cite{philips,WLEE2009,ORBITALORDERPRL}, low-frequency spin dynamics\cite{fernandes2012manifestations}, or, more recently, scenarios that incorporate both effects\cite{liangdagotto}.  Regardless of the primary driver of the electronic nematicity, a second debate has focused on the relationship between impurity scattering/in-plane defects and the origin of the nematic response.  This second debate is rooted in whether the dopant atoms themselves introduce anisotropic scattering effects\cite{UCHIDATRANSPORT,UCHIDAOPTICAL} that bias bulk measurements (such as charge transport studies) or whether the electronic anisotropy stems directly from a Fermi surface instability that is simply tuned via charge-doping\cite{kuo2013effect,ELASTORESISTIVITYKUO,fisher2011plane,Duszaanisotropic}.

Specifically, the bilayer pnictide system Ba(Fe$_{1-x}$Co$_{x}$)$_2$As$_2$ has provided a well-studied platform for exploring these scenarios.  In seminal charge transport studies, data showed that in-plane transport anisotropy surprisingly persisted well above the nominal tetragonal-to-orthorhombic transition temperature and the extent of this high temperature transport anisotropy evolved as a function of electron-doping\cite{chu2012divergent,kuomagnetoelastic,Blombergtensile,kuo2013effect}.  Subsequent studies, however, reported that post-growth annealing and alternative means of doping dramatically dampen this anisotropy\cite{UCHIDATRANSPORT,Yusongprb,ishida2013effect}, suggesting the dominant role of an anisotropic scattering mechanism driven by in-plane dopant impurities.  Adding to the debate, recent results have shown that, above the nominal T$_S$, strain-induced anisotropy is independent of relative levels of disorder in samples with similar antiferromagnetic (AF) ordering temperatures ($T_N$'s)\cite{kuo2013effect}.  Strain-induced anisotropy in this high-temperature, paramagnetic regime is widely interpreted as directly resulting from incipient nematic order; however direct measurements of the strain-induced response of correlated magnetic order and its evolution upon doping in this regime are notably lacking.

In this paper, we present neutron scattering measurements exploring the evolution of antiferromagnetism under applied uniaxial pressure as electron-doping is tuned in the bilayer iron pnictide compound Ba(Fe$_{1-x}$Co$_{x}$)$_2$As$_2$ (Co-doped Ba-122) with $x$ =0, 0.015, 0.030, and 0.040.   Our results map the response of long-range AF order to uniaxial strain and show that the strain-induced thermal shift in the onset of AF order is surprisingly insensitive to Co-doping in absolute terms.  As a percentage of the zero-strain $T_{N}$, however, the induced AF response increases with increased Co-doping---a result that challenges existing models of spin-lattice coupling in this compound.  We also demonstrate unambiguously that, on the first order side of the tricritical point in the magnetostructural phase diagram Co-doped Ba-122, the onset of orthorhombicity and AF order are decoupled under the application of sufficient uniaxial pressure and that, similar to measurements of the lattice strain susceptibility, the AF order parameter's response to strain is inherent to strains along the orthorhombic in-plane axes. Our results stand to provide valuable constraints on models of spin-lattice coupling in this compound as well as its relevance in the proposed nematic order parameter of this compound.

\section{Experimental Details}
All the crystals used in this study were prepared by standard self-flux techniques\cite{fluxmethodathena} and concentrations reported were determined via energy dispersive x-ray spectroscopy (EDS). The samples were used as-grown rather than being annealed for long periods of time.  We do not believe that any of the results in this study are affected by employing as-grown samples.  Thin plate-like crystals were cut with facets either along the in-plane [1,0,0] or the [1,1,0] orthorhombic axes and mounted within a pressure vise.  For the purposes of this paper, we will label all wave vectors $(H, K, L)$ using reciprocal lattice units [r.l.u.] where $\left|\textbf{Q}\right|$ [$\AA^{-1}$]$=\sqrt{(\frac{2\pi H}{a})^2+(\frac{2\pi K}{b})^2+(\frac{2\pi L}{c})^2}$.  $(H, K, L)$ vectors are given in the orthorhombic setting with $a\approx b \approx 5.60 \AA$ and $c\approx 13.02 \AA$ for the parent system.  The final pressure applied to samples loaded within the vise was determined via the compression of a Belleville washer in line with the piston and care was taken to apply approximately the same level of pressure to every sample studied.  We estimate that the initial loaded pressure varied no more than $10\%$ between samples with a nominal target pressure of $P=2$ MPa.  Pressure was applied at room temperature and the sample was loaded within a closed-cycle refrigerator with He-exchange gas.  

Data were collected first in the pressurized state and then in the zero pressure state (with the exception of the pressure applied parallel to the [1, 1, 0] axis in parent BaFe$_2$As$_2$ where only one pressure was measured).  This was done for consistency; however we found that the effects of uniaxial pressure we report to be reversible after several cycles of applying and releasing pressure.  The variance of the size of crystals studied spanned dimensions with length (parallel to the uniaxial pressure) $l=6\pm1.5$ mm, width (perpendicular to the pressurizing piston at the point of contact) $w=3\pm0.5$ mm, and thickness (uniform along the length of the sample) $t=0.23\pm0.05$ mm. The mosaic of each crystal studied was less than $0.5^{\circ}$ in the strain free case and remained unchanged within resolution once strained for all samples reported here.  Through initially aligning the pressurizing axis of the crystal vise perpendicular to the scattering plane, we were able to determine how far away from the nominal (0, K, 0) axis pressure was applied to the sample.  In all cases, the pressure was applied within under $\approx3 ^{\circ}$ of deviation from the (0, K, 0) axis. Once pressurized, the piston was locked into place via a setscrew to approximate uniform pressure while cooling.     

Neutron scattering measurements were carried out on the C5 triple-axis spectrometer ($x=0$) and N5 triple axis spectrometers ($x=0.015$ and $x=0.030$) at the Canadian Neutron Beam Center, Chalkriver Canada. Measurments were also performed on the HB-1 triple-axis spectrometer ($x=0.040$) in the High Flux Isotope Reactor (HFIR) at Oak Ridge National Laboratory.  Experiments on N5 and C5 were performed with a pyrolitic graphite (PG) monochromator and analyzer ($E_i=14.5$ meV) with a PG filter placed after the sample and collimations of $30^\prime- 60^\prime-sample-33^\prime-144^\prime$. The HB-1 setup consisted of a PG monochromator ($E_i=13.5$ meV), PG analyzer, two PG filters before the sample, and collimations of $48^\prime- 80^\prime-sample-80^\prime-240^\prime$.  For measurements with pressure applied parallel to the orthorhombic in-plane axes, samples were aligned within the $[H, 0, L]$ scattering plane, and the applied compressive pressure defined the short, $b$-axis which was oriented out of the scattering plane.

Finally, it is worth briefly describing one element of this manuscript's nomenclature:  In our experiments, we identify the onset of the structural distortion ($T_S$) via radial scans through the nuclear \textbf{Q}=(2, 0, 0) reflection.  Before pressure is applied, the width of this reflection changes substantially as the sample distorts into the orthorhombic phase and structural twin domains develop.  Here, we have previously shown that the temperature evolution of this peak width is a good approximation to the structural order parameter.  The center of the (2, 0, 0) peak however corresponds to a domain-weighted average lattice parameter, and this lattice value also shifts as the system distorts through $T_S$ due to the inequivalent expansion/contraction of the in-plane $a$/$b$-axes.  We simply label this value as the ``$a$-axis'' lattice constant since it is the apparent value in our scattering experiments.  For a fully detwinned sample (one in which the width of the (2, 0, 0) no longer changes through $T_S$), the quoted $a$-axis lattice constant is exact; however this transitions back to a domain weighted average under different levels of twinning.  For the purposes of our studies, we simply utilize the temperature evolution of the (2, 0, 0) reflection's width and effective lattice constant to resolve where the onset of $T_S$ occurs within resolution.

\begin{center}
\begin{figure*}[th]
\centering
\includegraphics[scale=.65]{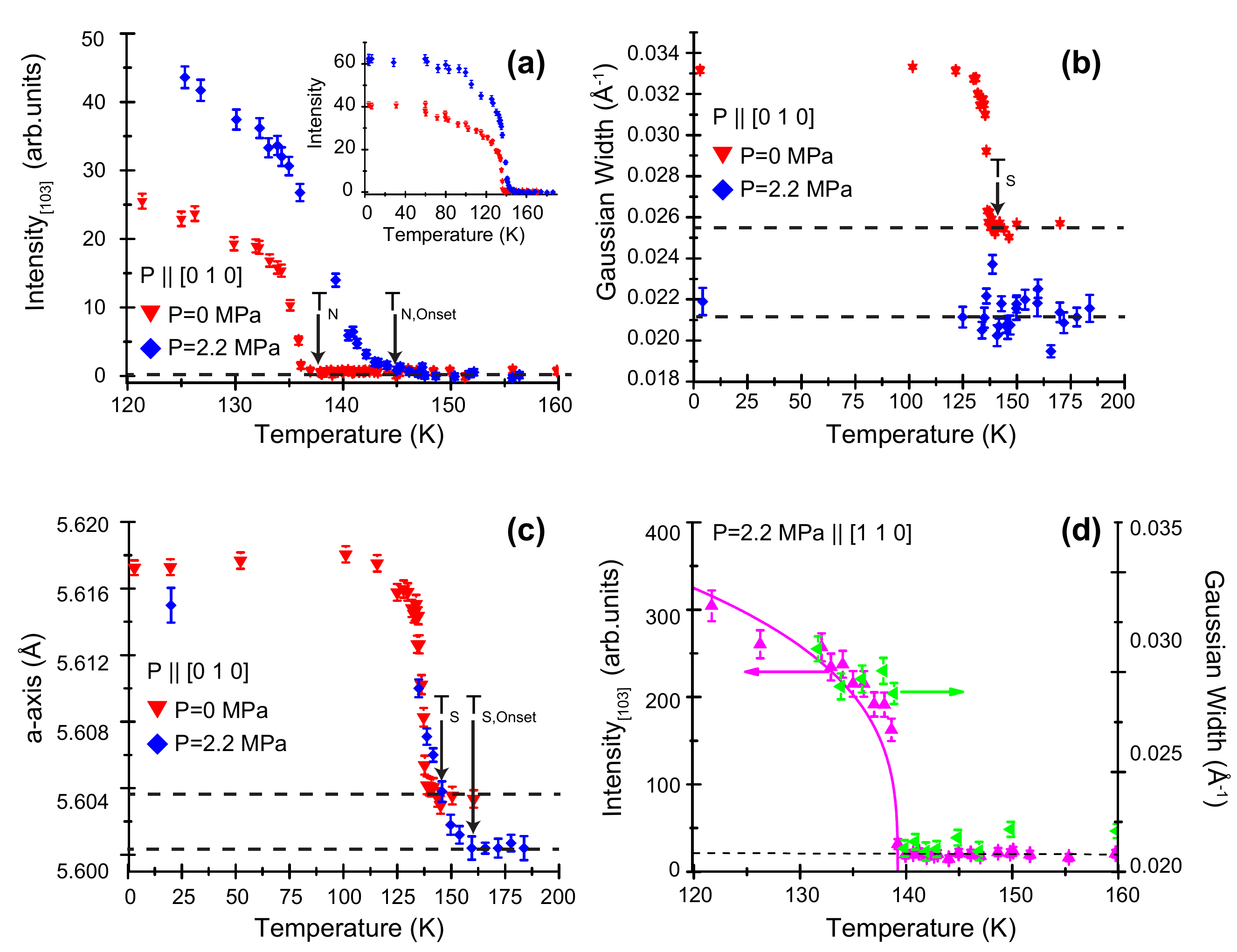}
\caption{Neutron scattering data collected on the BaFe$_{2}$As$_{2}$ (x=0) sample.  (a) Raw data showing the square of the magnetic order parameter collected at the (1, 0, 3) AF Bragg peak as a function of temperature.  Red triangles and blue diamonds show the evolution of AF order for 0 MPa and 2.2 MPa of uniaxial pressure applied respectively (b)  Temperature evolution of the fit Gaussian width of the \textbf{Q}=(2, 0, 0) reflection for both 0 MPa (red triangles) and 2.2 MPa (blue diamonds). (c) Temperature evolution of the effective $a$-axis for both 0 MPa and 2.2 MPa uniaxial pressure. (d) Raw data showing the magnetic order parameter squared collected at the (1, 0, 3) reflection and the Gaussian width of the nuclear (2, 0, 0) reflection plotted as a function of temperature.  These data were instead collected with 2.2 MPa uniaxial pressure applied along the [1, 1, 0] direction.}

\end{figure*}
\end{center}

\section{Results}
\subsection{$T_{N}$ under uniaxial pressure in B\MakeLowercase{a}F\MakeLowercase{e}$_2$A\MakeLowercase{s}$_2$}
As the baseline for our study, we explored the response of parent Ba-122 material under higher uniaxial pressure than those reported previously \cite{dhital2012effect}.   Figure 1 (a) shows the intensity of the antiferromagnetic \textbf{Q}=(1, 0, 3) reflection with uniaxial pressure initially applied along the [0,1,0] and then later released.  Consistent with earlier results, there is a sizable shift in the onset temperature of long-range AF order under the application of uniaxial strain with a high temperature tail in the order parameter that extends to $T_{N,Onset}=145$ K.

The apparent intensity of the magnetic peak increases due to the detwinning effect of uniaxial pressure which rotates a higher volume fraction of moments into the scattering plane.  Naively, one would expect the apparent magnetic intensity to double once the sample is completely detwinned from statistically equivalent domain populations into a single domain crystal; however we only observe a $\approx50\%$ increase in the magnetic intensity.  This occurs despite the fact that the in-plane nuclear reflections no longer exhibit any broadening at the structural distortion as plotted in Fig. 1 (b).  Curiously, near identical behavior ($\approx47\%$ increase) was observed in our previous study which applied less than half of the pressure utilized in the current case.  Our earlier interpretation\cite{dhital2012effect} that this effect was due to remnant twinning---twins hiding within the resolution of the scattering measurement---seems to be inconsistent with this coincidence.  The current experiment was performed on a separate sample and under twice the applied pressure, and it seems an unlikely coincidence that this and our previous measurements would result in identical ratios of remnant twin domains in different partially detwinned samples.  Rather, if we assume that the moments remain oriented rigidly along the $a$-axis under the applied pressure, the ordered moment appears to decrease in magnitude under the application of uniaxial pressure. The potential origins of this effect will be revisited in Section IV of this paper.

Data in Figs. 1 (b) and (c) show that, as the width broadening of the in-plane reflection \textbf{Q}=(2, 0, 0) vanishes under [0, 1, 0]-oriented uniaxial pressure, the lattice distortion shifts upward in temperature.  Due to the presence of a symmetry breaking uniaxial strain field, the distortion temperature $T_S$ is no longer rigorously defined.  For the purposes of our current study, we will define $T_S$ under uniaxial pressure as the temperature at which the structural distortion becomes resolvable within the resolution of our scattering experiments  ($T_{S,Onset}$)---in other words, where the sample's phase transition is detectable above any subtle distortion induced via strain.  Similar to the shift in AF order, the onset of $T_S$ broadens into a high-temperature tail under uniaxial pressure; however, notably, the onset of this tail extends far beyond the onset of long-range AF order with $T_{S,Onset}=157$ K.  The split in the onsets of $T_S$ and $T_N$ is unambiguously clear as AF order is much easier to detect in our measurements than a subtle shift in lattice parameters and clearly demonstrates that the AF and structural order parameters decouple under strain.

\begin{figure}
\includegraphics[scale=.75]{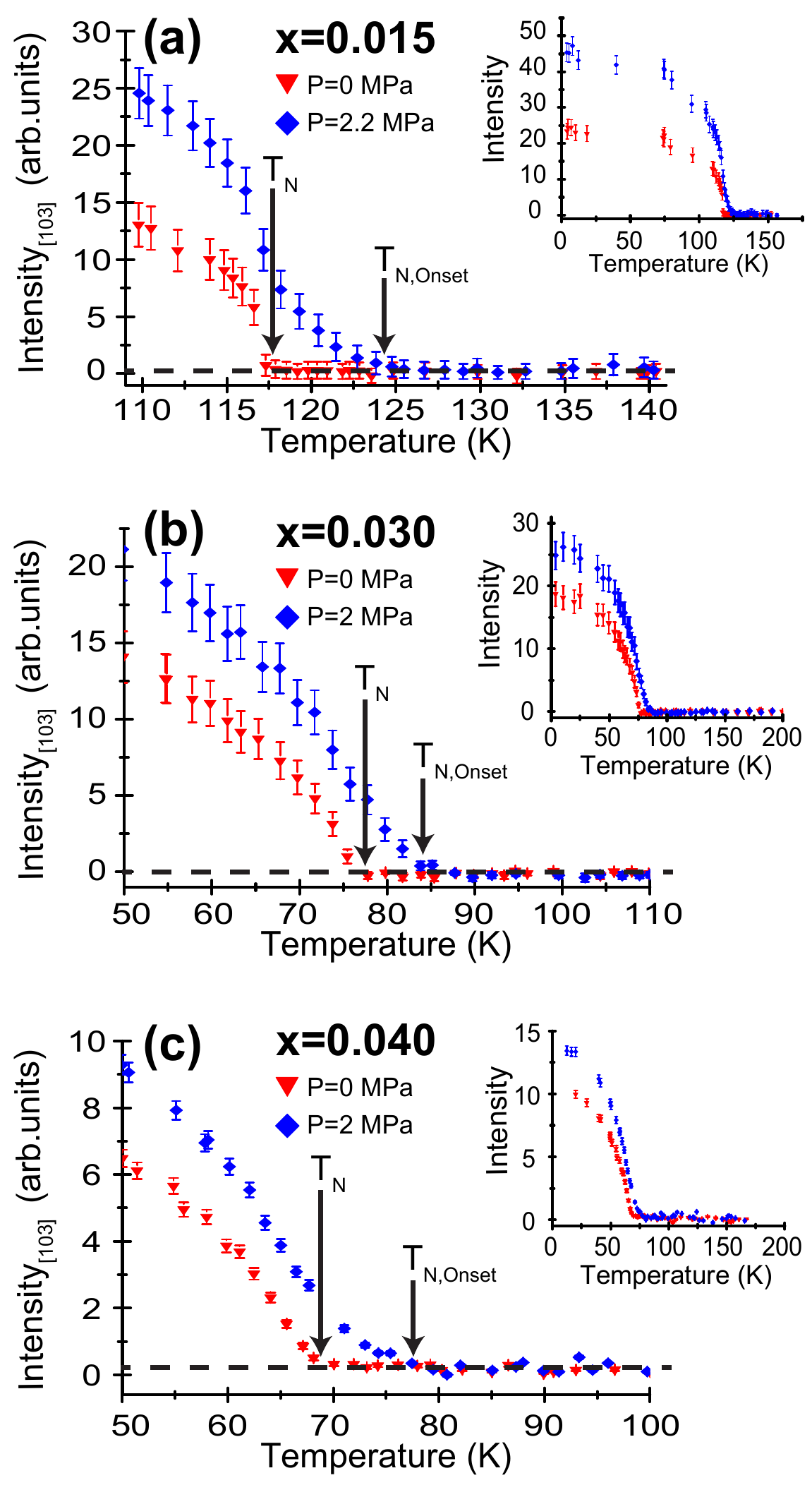}
\caption{Raw data showing the temperature evolution of the magnetic order parameter squared, collected at the (1, 0, 3) reflection under 0 MPa (red triangles) and 2 MPa (blue triangles) pressure applied along the [0, 1, 0] axis.  Data for the $x=$0.015, $x=$0.03, and $x=$0.04 samples are plotted in panels (a), (b), and (c) respectively.  Insets in each panel show an expanded view of thermal evolution of the order parameter.}
\end{figure}

Fig. 1 (d) plots both the magnetic and structural phase behaviors of a different BaFe$_2$As$_2$ sample with facets cut and comparable pressure applied along the in-plane [1, 1, 0] axis.  The intensity of the (1, 0, 3) AF peak turns on sharply at $T_N=139$ K with no strain-induced tail evident within the AF order parameter.  The structural distortion temperature $T_S$ as determined via the width change of the (2, 0, 0) nuclear reflection occurs simultaneous to $T_N$ (within 1K resolution), as expected for this nominal parent material.  We note here that the higher $T_N$ and $T_S$ of this crystal is an extrinsic sample dependence and falls within the range of typical values reported for Ba-122 which typically vary from 134 K to 140 K for as-grown crystals.  The absence of a high-temperature tail within the AF order parameter for the case of [1, 1, 0]-oriented pressure explicitly demonstrates that the strain-induced enhancement of AF order in this system stems solely from uniaxial strain fields oriented parallel to the in-plane orthorhombic axes and that radial stress effects within the sample do not affect the resulting phase behavior.  This directly parallels charge transport anisotropy effects\cite{fisher2011plane} and suggests that the enhancement of AF order and the large lattice response to uniaxial strain stem from the same susceptibility.  

\subsection{$T_{N}$ and $T_{S}$ under uniaxial pressure in B\MakeLowercase{a}(F\MakeLowercase{e$_{1-x}$}C\MakeLowercase{o$_{x}$})$_2$A\MakeLowercase{s}$_2$}

Turning now to Co-doped Ba-122 variants, similar measurements were performed with pressure applied along the [0, 1, 0]-axes of crystals on both the first-order and second-order sides of the tricritical point.  Data illustrating the response of the AF order parameter under comparable levels of uniaxial strain are plotted in Fig. 2.  Similar to the parent system under the application of uniaxial pressure, the onset of AF order shifts upward in temperature for all samples.  For the $x=0.015$ concentration, the first-order magnetic phase transition develops a prominent strain-induced tail, mirroring the parent phase behavior; however, unlike the parent material, the apparent magnetic intensity fully doubles in this sample under applied pressure.  This suggests a complete detwinning of the sample, although the behavior of the parent crystal (discussed earlier) suggests that the ordered moment may also evolve under uniaxial pressure.  For the two concentrations on the second-order side of the tricritical point (x=0.030 and x=0.040), applied pressure also manifests a similar high-temperature tail in the AF order which convolves with the power law behavior of the order parameter.

\begin{figure}[t]
\includegraphics[scale=.35]{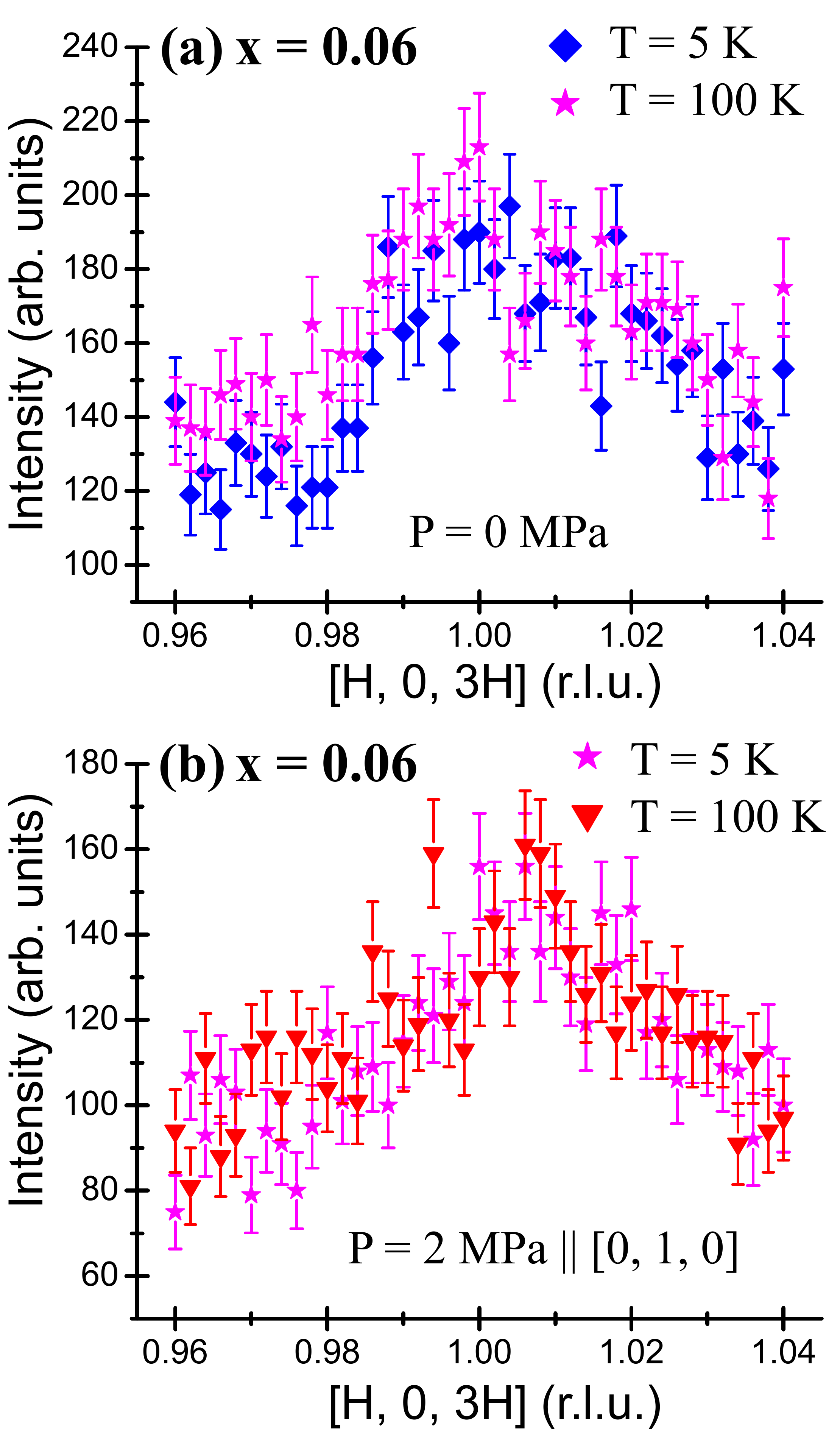}
\caption{Radial Q-scans through the (1, 0, 3) AF Bragg position both in zero pressure (a) and 2 MPa (b) for the $x=$0.06 Co-doped sample.  Data were collected at both 5 K and 100 K. }
\end{figure}

An additional sample with $x=$0.06 was also explored with neutron measurements.  This sample nominally lacks long-range AF order, and the goal here was to explore the possibility of inducing AF order under modest uniaxial pressure.  The data shown in Fig. 3 show that no ordered moment develops above 5K under the application of pressure, and the system remains paramagnetic within resolution.  This is consistent with the near vanishing of strain-induced transport anisotropy at this concentration\cite{fisher2011plane}.

Background subtracted radial scans collected within the tail of the strain-induced AF order parameter for each sample are plotted in Fig. 4.  The data here simply reinforce our earlier observation that the AF order induced by strain fields is long-range within the resolution of our measurements.  The difference in peak widths between Fig. 4 panels (a,b) and panel (c) arises from differing spectrometer resolutions stemming from the use of different instruments and collimations.   The experimental Bragg resolution (defined by the Gaussian full width at half-maximum of the resolution ellipsoid) for each measurement is illustrated as a central line in each panel for reference.  Low temperature radial scans deep within the AF ordered phase are also plotted for each sample in Fig. 4 panels (d-f).  These plots more clearly illustrate the gain in the apparent, saturated, long-range ordered AF moment under the application of uniaxial pressure.

\begin{figure}[t]
\includegraphics[scale=.45]{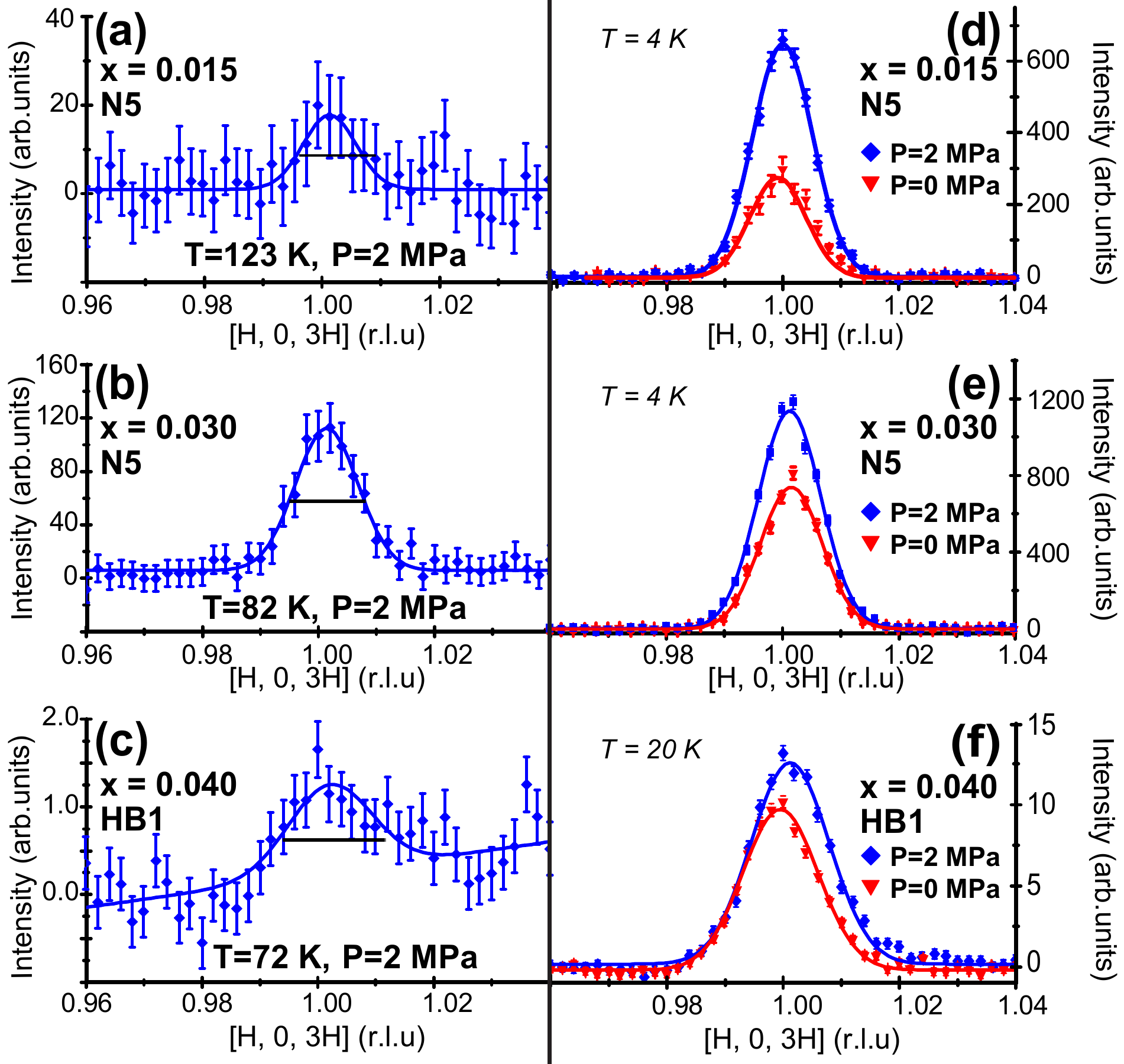}
\caption{Background subtracted radial Q-scans through the (1, 0, 3) AF Bragg position collected within the pressure-induced high-temperature AF tail of the order parameter for the $x = $0.015 (a), $x = $0.03 (b), $x = $0.04 (c) samples.  Data showing radial scans collected in the saturated region of the AF order parameter under both zero pressure and 2 MPa are plotted for $x = $0.015, $x = $0.030, and $x = $0.040 in panels (d), (e), and (f) respectively.}
\end{figure}

Data illustrating the structural distortion in these Co-doped samples are plotted in Fig. 5.  For the $x=$0.015 concentration, the in-plane nuclear (2, 0, 0) reflection distorts simultaneously with the onset of AF order under zero strain.  Under [0,1,0]-oriented pressure, this sample detwins within resolution, the onset of $T_S$ decouples from $T_{AF}$, and $T_S$ shifts substantially upward.  For the $x=$0.030 and $x=$0.040 concentrations, similar levels of uniaxial pressure also shift the onset of $T_S$ upward; however neither of these samples is appreciably detwinned under this same level of pressure.  This is consistent with higher Co-impurity concentrations pinning domain boundaries that subsequently require a higher strain field to bias through $T_S$.  While the precise pressure necessary to detwinn a sample is largely an extrinsic quantity, to the best of our knowledge there has been no systematic study reporting the evolution of the pressures necessary to detwinn Co-doped Ba-122 in similar quality samples.

An alternative means of analyzing the AF order parameter under strain is to fit the AF order parameters as simple power laws broadened by a distribution of ordering temperatures within the sample.  This broadening would potentially be due to an inhomogeneous strain field imposed across the crystal that nucleates AF order across a distribution of temperatures.  The presence of a severely inhomogeneous strain field would naively be unable to account for the sharp nuclear (2, 0, 0) peak throughout the structural phase transition in the parent and $x=0.015$ samples; however, if the volume fraction of high strain regions is small enough, it is conceivable the expected structural broadening may be diminished below experimental resolution.  In either case, fitting the magnetic order parameter to a Gaussian broadened power law behavior generates an alternative metric for assessing the influence of uniaxial strain on the development of AF order.

For each magnetically ordered sample, the data was fit to a power law of the form\cite{birgeneau1973spin}:
\begin{equation}
M^{2}(T)=\int_0^\infty (1-\frac{T}{T_{N}})^{2\beta}\frac{1}{\sqrt{2\pi}\sigma}\mathrm{e}^{(\frac{-(t_{\sigma}-T_{N})^2}{2\sigma^2})}\mathrm{d}t_{\sigma}
\end{equation}
Here $\beta$ is the critical exponent and $\sigma$ is the thermal width of the Gaussian distribution of $T_N$s within the sample.  It was assumed that the zero strain $\beta$ values remained unchanged upon the application of small levels of strain---this served to more reliably decouple the $\beta$ and $\sigma$ values as the system transitions into the second order regime.

\begin{center}
\begin{figure*}[th]
\centering
\includegraphics[scale=.6]{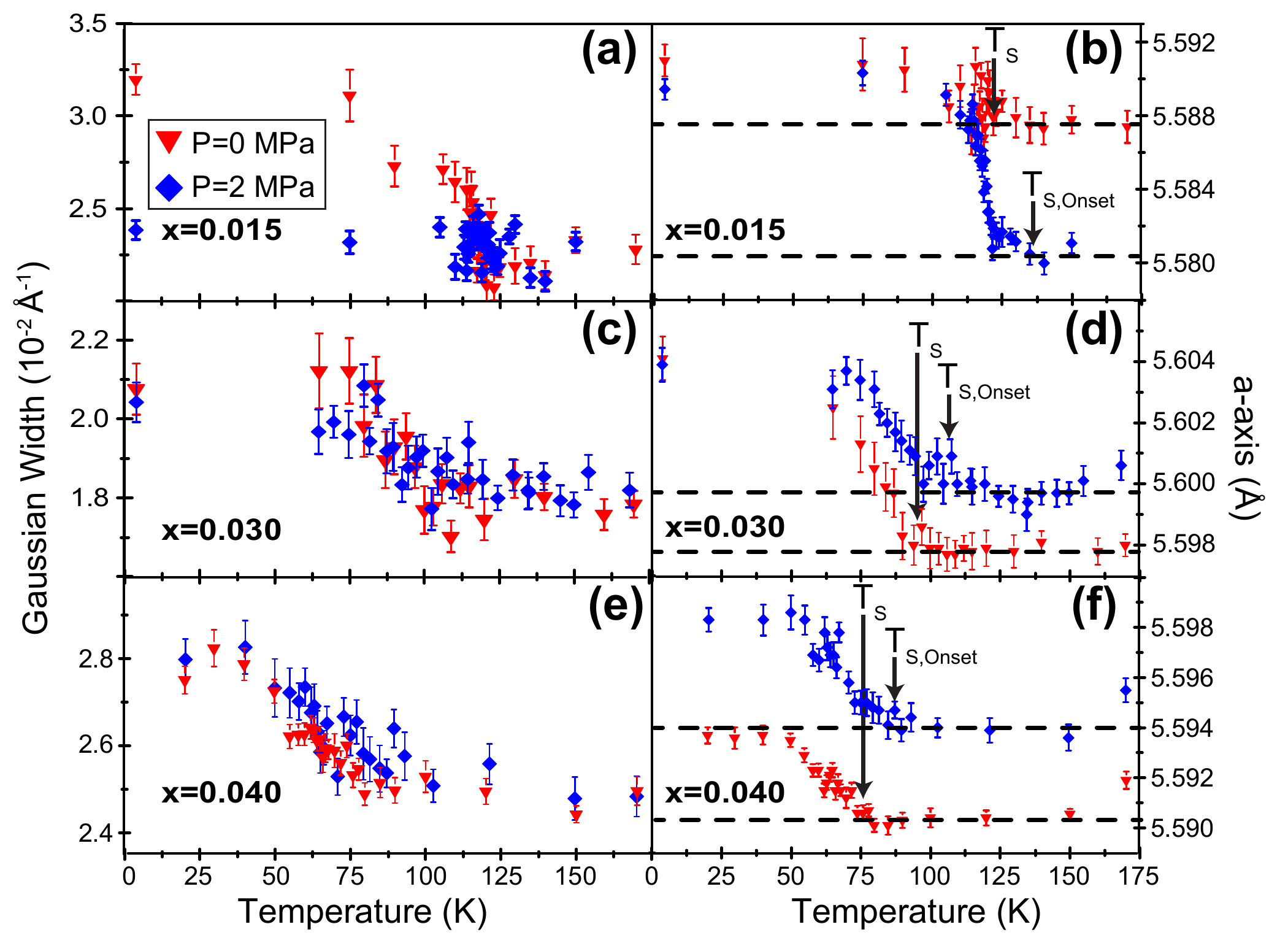}
\caption{Gaussian widths of the (2, 0, 0) nuclear reflection measured both under 0 MPa (red triangles) and 2 MPa (blue diamonds) uniaxial pressure for the (a) $x = $0.015, (c) $x = $0.030, and (e) $x = $0.040 samples.  The corresponding effective $a$-axis lattice parameters determined by the (2, 0, 0) peak position are plotted for these same samples in neighboring panels (b), (d), and (f) respectively. }
\end{figure*}
\end{center}

Fig. 6 shows the results of Gaussian-broadened fits to the square of the magnetic order parameters of the $x = $0, 0.015, 0.030, and 0.040 samples both with and without uniaxial pressure applied along the [0,1,0] axis.  Within error, the application of $\approx2$ MPa uniaxial pressure along the $b$-axis induces an increase of the effective Gaussian width of the distribution of $T_N$s by approximately 2 K for all samples.  This uniform increase in the distribution of $T_N$ effectively models the high temperature tail of the AF order parameter and is accompanied by an upward shift in the average $T_{N}$ of $\approx 3.5$ K (within error) for all samples excluding the $x = $0.015 sample.  This $x = $0.015 sample shows a minimal shift in its central $T_{N}$, potentially reflective of its closer proximity to the tricritical point at $x = $0.022.    The fit $\beta$ values using this fit method were consistent with earlier observations of a near two-dimensional Ising exponent on the first order side of the tricritical point with $\beta=0.110\pm0.02$ for $x=0$ and $\beta=0.14\pm0.02$ for $x  = 0.015$, which then transitions to $\beta=0.25\pm0.02$ for $x = $0.030 and $\beta=0.23\pm0.01$ for $x = $0.040 on the second order side of the tricritical point.  Technically, the AF transition in the $x = 0.015$ sample is weakly first order but the AF peak intensity is nevertheless effectively modeled by a rounded power law.  Ultimately, this alternative form of analyzing the magnetic order parameters under strain reveals a similar conclusion to the simple method identifying the AF onset temperature discussed previously---namely that the absolute thermal shift in the onset of long-range AF order under pressure is nearly Co-doping independent.

Finally, as a summary of the influence of uniaxial strain on the AF order parameter, Fig. 7 plots the doping-dependent shift in $T_{N}$ under the application of fixed uniaxial pressure.  The thermal shift is plotted via two different methods:  The first is shown in Fig. 7 (a) and shows the pressure-normalized shift in the onset of $T_{N,Onset}$ upon applying pressure along the [0, 1, 0] axis.  Specifically, the plotted quantity is $[\frac{dT_{N}}{P}]_{Onset}=\frac{T_{N,Onset}(P)-T_{N,Onset}(0)}{P}$. Here $T_{N,Onset}$ is determined empirically at the first temperature at which long-range AF order is observed above the background, and $P$ is the applied pressure.  The second panel, Fig 7 (b), plots the shift in the mean $T_{N}$ plus the increase in half-width at half-maximum of the modeled distribution of  $T_{N}$'s using the Gaussian-broadened power law fits plotted in Fig. 6.  Explicitly, we defined a quantity $T_{N,Avg}=T_{N}$+$\sigma\sqrt{2ln2}$ (values shown in Table I) to define the effective shift in the leading edge of the tail of the AF order parameter using this alternative metric and the corresponding relation $[\frac{dT_{N}}{P}]_{Avg}=\frac{T_{N,Avg}(P)-T_{N,Avg}(0)}{P}$ .  In both cases, the shift in the effective onset of long-range AF order under the application of modest uniaxial strain is seemingly independent of doping.

\begin{figure}[t]
\includegraphics[scale=.4]{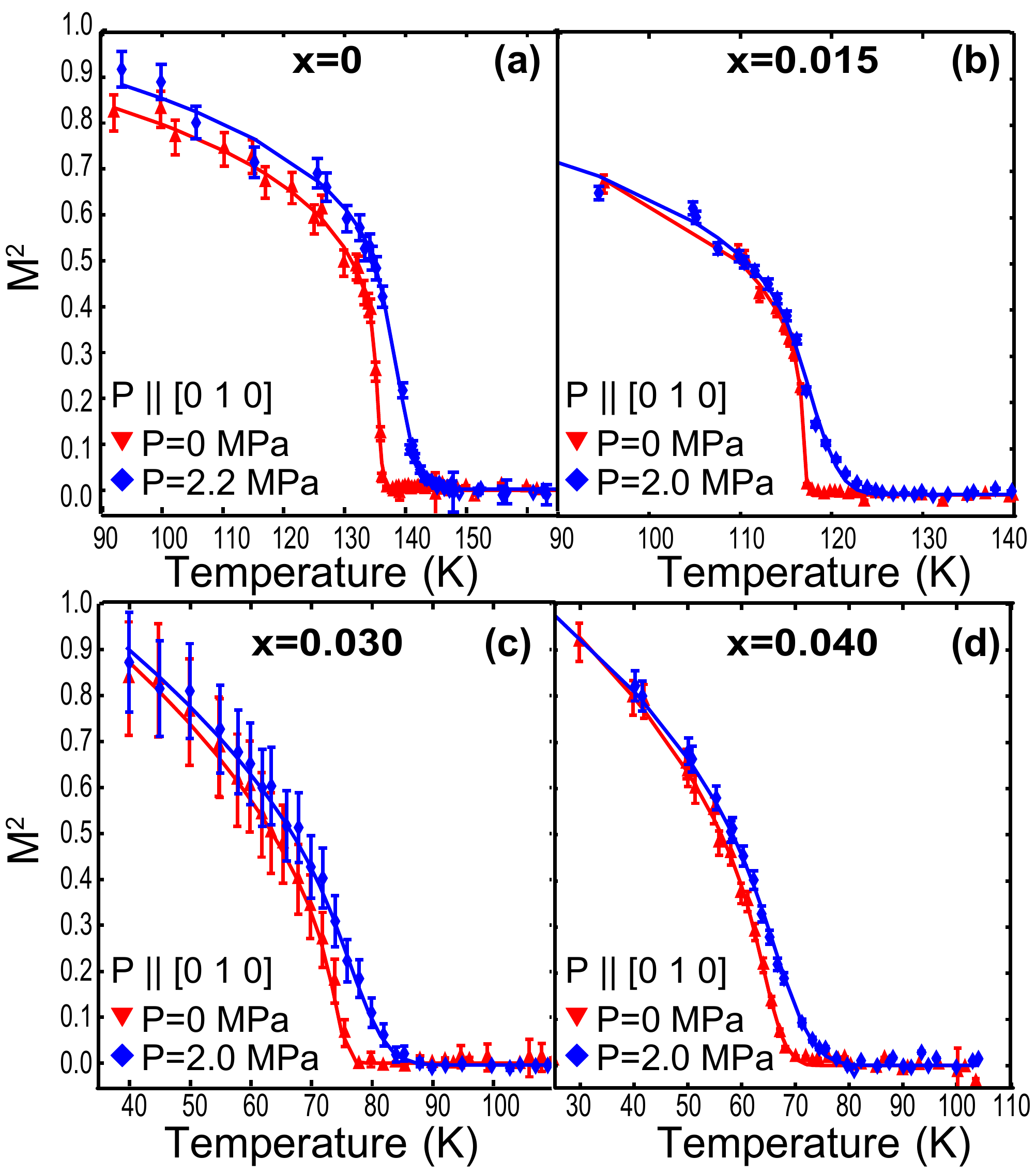}
\caption{The results of power law fits with a Gaussian distribution of $T_{N}$'s (as described in the text) are plotted in panels (a-d) for the $x=$0, 0.015, 0.030, and 0.040 samples respectively.  Fits are plotted for both 0 MPa (red triangles) and 2 MPa (blue diamonds) pressures.  The corresponding fit parameters are summarized in Table 1.}
\end{figure}

\begin{table}[]
\centering 
\begin{tabular}{c c c c c c} 
\hline\hline 
x & $T_{N}$(0 MPa) & $\sigma$(0 MPa) & $T_{N}$(2 MPa) & $\sigma$(2 MPa) & $2\beta$ \\[0.5ex]
\hline 
0.000 & 135.6$\pm$0.1 & 0.6$\pm$0.1 & 139.1$\pm$0.3 & 2.6$\pm$0.2 & 0.22$\pm$0.02 \\ 
0.015 & 117.1$\pm$0.1 & 0.23$\pm$0.06 & 118.4$\pm$0.3 & 2.5$\pm$0.2 & 0.27$\pm$0.02 \\
0.030 & 75.2$\pm$0.2 & 1.3$\pm$0.4 & 79.6$\pm$0.6 & 4.2$\pm$0.7 & 0.49$\pm$0.03\\
0.040 & 66.0$\pm$0.1 & 2.8$\pm$0.3 & 69.3$\pm$0.5 & 4.7$\pm$0.5 & 0.46$\pm$0.02 \\ [0.5ex] 
\hline 
\end{tabular}
\caption{Parameters for Gaussian-broadened power law fits of AF order parameters as described in the text.  Units for temperatures and Gaussian widths are in Kelvin.} 
\end{table}

\begin{figure}[t]
\includegraphics[scale=.65]{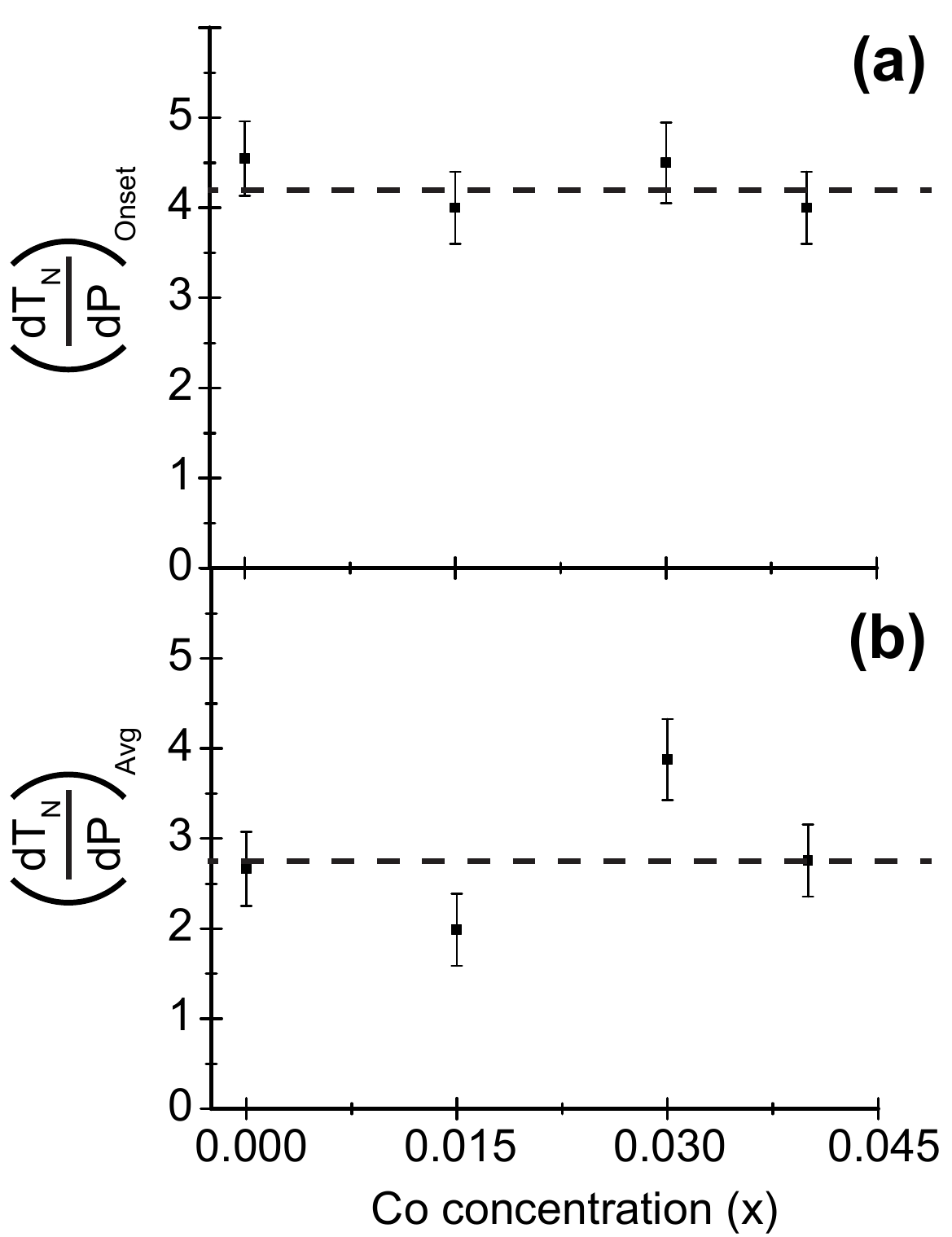}
\caption{The shift in the empirically observed onset of  $T_{N,Onset}$ normalized by the applied pressure is plotted as a function of doping in panel (a).  The doping dependent shift in the average $T_{N,Avg}$ determined by the fitting parameters of Table 1 and defined in the text is normalized by the applied pressure and plotted in (b). }
\end{figure}

\section{Discussion}
Our findings demonstrate a surprising insensitivity of the strain-induced shift in the onset of AF ordering as Co-impurities are introduced into Ba-122.  The absolute shift in the onset of AF order is seemingly independent of Co-concentration below 4$\%$, and the relative change therefore necessarily increases as a fraction of $T_N$ as the AF phase is weakened upon electron-doping.   This finding is seemingly at odds with previous phenomenological models which predict a decrease in the strain response of AF order as the structural and AF phase transitions are decoupled upon electron-doping.  For instance, a previous analysis based on Ginzberg-Landau treatment of the magnetoelastic coupling of the structural and magnetic order parameters predicts that the shift in T$_N$ under uniaxial stress ($\sigma$) should vary as  $\frac{\delta T_{N}}{\delta\sigma} \propto (T_S - T_N)^{-1}$.\cite{Cano}  Similarly, a minimal microscopic $J_1 - J_2 - J_z$ model with nearest neighbor biquadratic coupling\cite{Kivelson} suggests that the shift in $T_N$ should scale as $(T_S - T_N)^{- \gamma}$ with an exponent ${\gamma > 2}$. Recent Monte Carlo simulations simultaneously treating both the spin-lattice and orbital-lattice couplings within the Hamiltonian suggest that both are necessary to model the resulting phase diagram and nematic behavior in this system \cite{liangdagotto}.  This suggests that additional degrees of complexity such as modified orbital-lattice coupling upon electron-doping will likely need to be accounted for in future theoretical efforts to model AF order's response to symmetry breaking strain fields in Co-doped Ba-122.

Generally, in the presence of strain, the $C_4$ symmetry is broken, and we don't expect a sharp structural transition in our measurements. For the scenario in which orbital ordering drives the underlying lattice instabiltiy, the onset of the $C_4$ symmetry breaking structural orthorhombicity due to the applied strain-field lifts the degeneracy of the $d_{xz}$ and $d_{yz}$ orbitals and rounds off the orbital ordering transition. This enables a small but non-zero orbital imbalance at higher temperatures over the regime where the orthorhombic structure distortion is observed. This orbital configuration potentially promotes the AF SDW order and increases the transition temperature $T_N$ below which the required time-reversal symmetry breaking takes place.

Our data suggest a reduction in the AF ordered moment of parent BaFe$_2$As$_2$ under uniaxial pressure, which is consistent with recent predictions from \textit{ab initio} density functional theory (DFT) \cite{Tomic}.  Pressure-induced changes in pnictogen height are predicted to modify the resulting ordered moment as in-plane stress is applied; however we note here that the effect we observe occurs at significantly lower pressures than those modeled in Ref. 49.  It is difficult to completely preclude the effect of remnant twin domains changing the apparent moment value, but to provide an estimate, we can simply assume that the structural peaks serve as a reliable indicator for when the sample has been completely detwinned.  Using this assumption, the ordered moment has been reduced by $12\%$ relative to its stress-free value under the application of $\approx2$ MPa.  Upon doping a slight amount of Co-impurities however, this effect seems to diminish and the 1$\%$ Co-doped sample shows complete detwinning without an effective AF moment change under an similar level of pressure. Future measurements with higher momentum resolution will be required to unambiguously determine whether the moment is suppressed via strain within the parent system.

The decoupling of $T_S$ and $T_N$ under uniaxial pressure suggests that the magnetic order shifts upward in temperature as a secondary effect driven by the pressure-enhanced orthorhombicity of the lattice.  This decoupling occurs for both Co-concentrations measured below the magnetostructural tricritical point in the electronic phase diagram (x=0, x=0.015), and more generally the separation between $T_N$ and $T_S$ in all samples appears to depend on the magnitude of the applied uniaxial pressure.  From phenomenological models, the differing response of both $T_S$ and $T_N$ to strain can readily be explained via the magnetoelastic coupling constant which dampens the shift of AF order relative to the shifted temperature at which significant orthorhombic distortion sets in.  Another possible explanation may arise from a varying response of spin-lattice and spin-orbital coupling strengths which respectively tune the relative response of the AF and structural order parameters to applied pressure \cite{liangdagotto}.

\section{Conclusion}
We have shown that the application of uniaxial strain along the in-plane [0, 1, 0] axis of Co-doped Ba-122 induces an upward shift in the onset of AF order for all samples which possess AF order in the strain-free state.  The separation of the onset of AF order and significant orthorhombicity evolves as a function of applied uniaxial pressure, and for samples on the first order side of the tricritical point, the onsets of $T_S$ and $T_N$ decouple.  Under the application of a near identical level of uniaxial pressure, the shift in $T_N$ as a function of Co-doping is seemingly uniform in absolute terms; however it correspondingly diverges as a fraction of $T_N$ as Co-doping suppresses AF order.  This varies from the expectations of existing theoretical models of the magnetostructural phase behavior in this material, suggesting that added effects such as modified orbital-lattice coupling as a function of Co-doping should be accounted for.  Our results will hopefully stimulate further theoretical efforts to fully explain the complex coupling between AF order and the orthorhombic/nematic phase behavior in this class of materials.

\acknowledgments{
The work at BC was supported by NSF CAREER Award DMR-1056625 (S.D.W) and DOE DE-FG02-99ER45747 (Z.W.).  The work at LBL was supported by the Director, Office of Science, Office of Basic Energy Sciences, US Department of Energy, under Contract No. DE-AC02-05CH11231.  Research conducted at ORNL's High Flux Isotope Reactor was sponsored by the Scientific User Facilities Division, Office of Basic Energy Sciences, US Department of Energy.}

\bibliography{CoDopedBa122UniaxialStrain_v5}

\end{document}